\documentclass{article}

\usepackage{arxiv}

\usepackage{cite}
\usepackage{amsmath,latexsym,amssymb,amsfonts}
\usepackage{algorithmic}
\usepackage{graphicx}
\usepackage{textcomp}
\usepackage{xcolor}
\usepackage{multirow}
\usepackage[font=footnotesize]{subfig}
\usepackage{subfig}
\usepackage{graphicx}
\usepackage{balance}
\usepackage{array}
\usepackage{amssymb}
\usepackage{hyperref}
\usepackage{mathtools}
\usepackage{tikz}
\usepackage{url}
\usepackage{amsmath}
\usepackage{makecell}
\usepackage[algo2e,ruled,vlined,linesnumbered]{algorithm2e}
\usepackage[nodisplayskipstretch]{setspace}

\newtheorem{theorem}{Theorem}[section]

\newtheorem{mydef}{Definition}[section]
\usepackage{pgfplots}
\usepackage{siunitx}

\title{Synthesis of Feedback Controller for Nonlinear Control Systems with Optimal Region of Attraction}

\author{
  Ayan Chakraborty \\
  Department of Computer Science\\
 Indian Institute of Technology Kanpur\\
   Kanpur,Uttar Pradesh : 208016\\
   \texttt{ayancha@cse.iitk.ac.in} \\
   \And
  Indranil Saha \\
  Department of Computer Science\\
   Indian Institute of Technology Kanpur\\
   Kanpur,Uttar Pradesh : 208016\\
  \texttt{isaha@cse.iitk.ac.in} \\
}

\begin{document}
\maketitle

\begin{abstract}
We propose a framework for synthesizing a feedback control policy that maximizes the region of attraction (ROA) of a closed-loop nonlinear dynamical system. Our synthesis technique relies on stochastic optimization, which involves computation of an objective function capturing the ROA for a feedback control law. We employ a machine learning technique based on deep neural network to estimate the ROA for a given feedback controller. Overall, our technique is capable of synthesizing a controller co-optimizing traditional control objectives like LQR cost together with ROA. We demonstrate the efficacy of our technique through exhaustive experiments carried out on various nonlinear  systems.
\end{abstract}

\section{Introduction}
\label{sec-introduction}

 With growing complexity of cyber-physical systems and robotics , guarantees in  performing certain tasks successfully are required  while an agent interacts with the environment. This requirement  is an essential feature in automation. For example, multiple agents interacting with the environment should ensure to achieve the specified objectives. An ambitious goal is to design an autonomous system that would synthesize controller(s) for the system regardless of how the environment behaves. 
 This is a fundamental problem in AI and Computer Science that has been extensively studied under different titles by control communities~\cite{1a,1b,1c}. 

Modern cyber-physical systems rely heavily on the efficacy of the feedback controllers. The efficacy of a feedback controller is generally measured by its capability of keeping the system stable at an equilibrium point and making it follow a reference trajectory precisely. 

Moreover, for real-world safety-critical systems, we need to avoid certain states from which the system cannot recover to ensure safe operation.
Thus, for safety-critical systems, another important property is the \emph{Region of Attraction} (ROA)~\cite{Khalil02} which represents the region of the state-space from where if the system initiates its operation, it is guaranteed that the system will remain inside the region during its entire operation and eventually reach an equilibrium state. Though the feedback controller synthesis problem for stability and trajectory tracking has been widely studied~\cite{astrom2010feedback,Khalil02}, the feedback controller synthesis problem for maximizing the ROA of a nonlinear dynamical system has not received such attention. 

The major challenge for synthesizing a controller with the largest ROA is that the ROA cannot be represented as a closed-form function of the components involved in the dynamics of the closed-loop system. Thus, gradient based optimization procedures are not applicable to synthesis of a controller with the size of the ROA of the closed-loop system as the objective function. Evolutionary algorithms have been used to solve controller synthesis problem with complex objectives in the past (e.g.~\cite{MajumdarSZ12}). However, such methods require a true measure of the objective function for a given candidate solution.
For a nonlinear dynamical system, \emph{Lyapunov functions} are the most convenient tools for safety certification and hence \emph{ROA} estimations~\cite{VannelliV85,SilvaT05}.
Even though searching such function analytically is not a straight forward task but can be identified efficiently via a  semi definite program~\cite{Parrilo00,BV2014} , or using SOS polynomial methods~\cite{HenrionK14}. Some other methods to obtain  \emph{ROA} includes volume over system trajectories, sampling based approaches~\cite{BobitiL16} and so on.
The state-of-the-art technique for estimating the ROA for a nonlinear dynamical system is the Sum-of-Square (SOS) procedure~\cite{HenrionK14}, which is based on finding a Lyapunov function of a pre-decided from through Semidefinite Programming (SDP)~\cite{Parrilo00,BV2014}. However, the major disadvantage of using the SOS procedure is that it is often not clear how close the estimated ROA is to the actual ROA of the system. Thus, given two controllers $K_1$ and $K_2$, if the estimated ROA for $K_1$ is larger than that for $K_2$, it is not possible to decide with certainty that the true ROA for the closed-loop system with controller $K_1$ is also larger than the true ROA for the closed-loop system with controller $K_2$.

Recently, Berkenkamp et al.~\cite{BerkenkampMS016,RichardsB018} have proposed a deep neural network based methodology to compute a close approximation of the ROA of a nonlinear dynamical system.  
It combines ideas of Gaussian Process (GP) learning~\cite{Seeger04} to approximate the model uncertainties and Lyapunov stability theory~\cite{Khalil02} to estimate the safe operating region. The key limitation is however, given a controller it only provides an algorithm to compute its ROA. Given a nonlinear dynamical system, a radical question is how do we synthesize a feedback controller that helps the dynamical system to achieve the best possible ROA. This also leads to one of the most pivotal decision of parameters tuning while designing a controller for a dynamical systems.

 Some earlier techniques involved \emph{quantization error} criteria to synthesize feedback controller~\cite{RichardsB018}. Off-late many of the developments happened by using deep reinforcement learning (Deep RL) methods to learn continuous control policies for dynamical systems~\cite{Khalil02}. The goal is generally to synthesize a feedback controller that has a very good trajectory tracking performance, which is captured in the form of so called \emph{LQR cost}~\cite{LewisVS12}. However, the controller having the best LQR cost may not achieve the best ROA for 
 the dynamical system.

In this paper we address these limitations and make the following contributions: We present a novel method for synthesizing an optimal controller that provides a very good tracking performance together with achieving the best possible ROA for the closed-loop system. Our technique involves a stochastic optimization technique to solve the best ROA feedback controller synthesis problem, keeping also in mind  the LQR cost as a performance criteria. As a specific technique, we  use the \emph{Particle Swarm Optimization} (PSO)~\cite{KennedyE95} method to find the  best policies.

We have developed a software tool by implementing the proposed methodology in Matlab and applied the tool for synthesizing feedback controllers for two different models of inverted pendulum, a vehicle steering system, and an aircraft pitch control system. For all the systems, we have been able to synthesize a feedback controller that improves the ROA with respect to the LQR controller significantly, without compromising much on the LQR cost. For a pendulum system, we demonstrate that the controller synthesized by our technique can stabilize the pendulum at the vertical position from an initial angle from where the LQR controller does not succeed to stabilize the system.

\section{Problem}

\label{sec-problem}

\subsection{Preliminaries}
\label{sec-prelim}
We use $\mathbb{N}$, $\mathbb{R}$,  $\mathbb{R}^+$, and $\mathbb{R}_0^+$ to denote the set of all natural numbers, the set of all real numbers, the set of all positive real numbers, and the set of all non-negative real numbers respectively. We use $\mathbb{R}^m$ to denote the $m$ dimensional real space.


\smallskip
\noindent
\textit{Dynamical System and Control Policy.}
We consider a nonlinear, deterministic dynamical system:
\begin{align}
\label{e1}
   \dot{\xi}(t) = f(\xi(t),\upsilon(t)),
\end{align}
where $\xi(t) \in \mathcal{S} \subset \mathbb{R}^n$ and $\upsilon(t) \in \mathcal{U} \subset \mathbb{R}^m$ are the states 
and control inputs at time $t \in \mathbb{R}_0^+$. The system is controlled by a feedback policy $\pi : \mathcal{S} \longrightarrow \mathcal{U}$, 
thus the closed loop dynamics is given  by $\dot{\xi}(t) = f(\xi(t),\pi(\xi(t)))$. 

The nonlinear dynamical system in~\eqref{e1} can be approximated into a \textit{linear} control system captured by linear differential equation:
\begin{equation}
\dot\xi(t)=A\xi(t)+B\upsilon(t),
\label{control_system}
\end{equation}
where 
$A$, $B$ are matrices of appropriate dimensions. 
The curve $\xi:\mathbb{R}_0^+ \mapsto \mathbb{R}^n$ is a \textit{trajectory} of (\ref{control_system}) if there exist a curve $\upsilon:\mathbb{R}_0^+\mapsto \mathbb{R}^m$ such that the time derivative of $\xi$ satisfies (\ref{control_system}). 
To keep the disposition simple, we assume that the system is fully observable. However, our technique can be seamlessly extended to partially observable system by introducing observers suitably.

For the sake of computer-based implementation of the control system, we consider the discrete-time version of (\ref{control_system}), as follows:
\begin{equation}
x[r+1]=A_\tau x[r]+B_\tau u[r]
\label{control_system1}
\end{equation}
where the matrices $A_\tau$ and $B_\tau$ are given by:
\begin{align}\nonumber
A_\tau=&\textsf{e}^{A\tau},~~~B_\tau=\int_{r\tau}^{(r+1)\tau}\textsf{e}^{A(\tau-t)}Bdt,
\end{align}
and $\tau$ is the sampling time. The function $\textsf{e}^{At}$, for any $t\in\mathbb{R}_0^+$, denotes the matrix function defined by the convergent series:
\begin{equation}\nonumber
\textsf{e}^{At}=I_{n}+At+\frac{1}{2!}A^2t^2+\frac{1}{3!}A^3t^3+\cdots,
\end{equation}
where \textsf{e} is Euler's constant. 
The signals $x$ and $u$ describe the exact value of the signals $\xi$ and $\upsilon$ respectively, at the sampling instants $0,\tau,2\tau,3\tau,\ldots$.
Mathematically, we have:
$$x[r]=\xi(r\tau),~u[r]=\upsilon(r\tau).$$
In this paper, we consider that the control policy $\pi$ is obtained by using proportional control, where a proportional gain matrix $K \in \mathbb{R}^{m\times n}$ is used to obtain the control policy $\pi : \mathcal{S} \longrightarrow \mathcal{U}$ as follows: $u[r]=-Kx[r]$, where $x[r] \in \mathcal{S}$ and $u[r] \in \mathcal{U}$. 
We call $K$ a \emph{feedback controller} and denote the space of all feedback controllers by $\mathcal{K}$.
For a feedback controller $K$, the closed loop dynamics is denoted by $f_K$.

\smallskip
\noindent
\textit{Linear Quadratic Regulator (LQR) Controller.}
There exist many algorithms for synthesizing a feedback controller for a dynamical system. An LQR controller is a popular feedback controller that is synthesized by keeping a balance between the trajectory tracking performance and energy spent in generating the control signals.

\begin{mydef}(\textit{LQR Cost}~\cite{astrom2010feedback})
Given a discrete time state-space model in~\eqref{control_system1},
the LQR cost is given by the following quadratic cost function.
 \begin{equation}
     J_{LQR} = \sum_{n=1}^{\infty}\left( x[r]^\textsf{T} Q x[r] + u[r]^\textsf{T} R u[r] \right)
\label{LQR_cost}
 \end{equation}
 for some given positive definite matrices $Q$ and $R$ of suitable dimensions.
\end{mydef}

The controller that minimizes the cost function in \eqref{LQR_cost} for a given dynamical system is called the \emph{LQR controller} and is denoted by $K_{LQR}$.

\smallskip
\noindent
\textit{Region of Attraction (ROA).}
We now define \emph{region of attraction} (ROA) which indicates a safe-operating region for a given controller.

\begin{mydef}(\textit{ROA}~\cite{astrom2010feedback})
Let $\xi_0$ be an asymptotically stable equilibrium point (state) for a given closed-loop dynamical system with state space $\mathcal{S}$. A subset $\mathcal{R}$ of state space $\mathcal{S}$ is called the \emph{region of attraction} if all the trajectories initiating from a state in $\mathcal{R}$ always remain in the region $\mathcal{R}$ and converge to $\xi_0$ as $t \to \infty$.
\end{mydef}

For a given controller $K \in \mathcal{K}$, we denote the largest possible ROA by $\mathcal{R}_K$. For a given open-loop dynamical system, we denote the feedback controller that maximizes the size of the ROA by $K_{max}$.


\subsection{Problem Definition}
Though both the LQR cost and the ROA for a controller are important performance criteria, the following example illustrates that they are unrelated. 
The LQR controller may not provide the best possible ROA among all feedback controllers. There may exist a feedback controller whose LQR cost is worse than the LQR controller but has an ROA which is significantly larger than the ROA of the LQR controller.

Keeping the above fact in mind, we define the controller synthesis problem as a multi-objective optimization problem where we attempt to synthesize a feedback controller for a dynamical system by co-optimizing both the LQR cost and the ROA. In the synthesis process, our goal is to find a controller that minimizes the LQR cost and maximizes the ROA.  The optimization problem is given by

\begin{equation}
\underset{K \in \mathcal{K}}{\text{minimize}} \ \  \omega_1 \times J_{LQR}(K) - \omega_2 \times \mathcal{R}_{K}
\label{obj_func}
\end{equation}
where $\omega_1, \omega_2 \in \mathbb{R^+}$ are two weights to be chosen by the user to capture the importance of the two objectives for a given application.

We denote the controller synthesized by optimizing the above cost function by $K_{\mathcal{O}}$.

\subsection{Illustrative Example.}
\label{sec-example}

\begin{table}[t]
\small
\centering
\begin{tabular}{|c|c|c|}
\hline
\thead{\textbf{\texttt{Policy }}\\w.r.t. $K_{LQR}$} &\thead{ \textbf{ $\%$  \texttt{Increase}}\\ \textbf{\texttt{in LQR cost}}} & \thead{\textbf{ $\%$ \texttt{Increase}}\\ \textbf{ \texttt{ in ROA}}} \\
\hline
$K_{max}$ & $6.1$ & $33.5$\\\hline
$K_{\mathcal{O}}$ & $4.02$ & $32.67$\\\hline 
\end{tabular}
\caption{Performance Comparisons of policies w.r.t \emph{LQR}-controller}
\label{T1}
\end{table}

Let us describe the problem introduced in the previous section with an example of an \emph{inverted pendulum}. The system is governed by the second order differential equation :
\begin{align*}
    &\ddot{\varphi} = \frac{g}{\ell} \sin \varphi -\frac{\mu}{\mathfrak{I}} \dot{\varphi} + \frac{1}{\mathfrak{I}} u\\
    &\varphi(0) = c \\
    &\dot{\varphi}(0) =0
\end{align*}
where $\xi := [\varphi,\dot{\varphi}]$ is the state vector,
$\varphi$ is the angle from the upright equilibrium position, $\mathfrak{I}$ is the moment of inertia of the pendulum, $l$ is the length of the pendulum, $\mu$ is a frictional coefficient, and $g$ is the gravitational constant. Given this system, we derive three different controllers by choosing the value of $\omega_1$ and $\omega_2$ in the objective function in Equation~\eqref{obj_func}: $K_{LQR}$ (LQR controller; $\omega_1 =  1$, $\omega_2 = 0$), $K_{max}$ (the controller with the optimal ROA; $\omega_1 = 0$, $\omega_2 = 1$), and  $K_{\mathcal{O}}$ (the optimal controller that is obtained by co-optimizing the LQR cost and the ROA; $\omega_1 > 0$, $\omega_2 > 0$). 

\begin{figure}[t]
\centering
 \includegraphics[width=8cm, height=6cm]{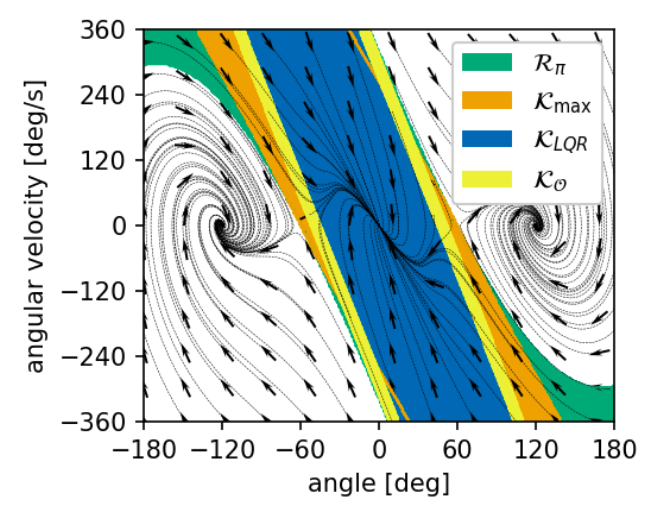}
 \caption{Comparisons of convergence to the safe set $\mathcal{R}_{\pi}$ w.r.t the controllers $K_{LQR},K_{max},\text{ and }K_{\mathcal{O}}$ }
 \label{F}
  \end{figure}

Figure~\ref{F} shows the ROAs for the three different controllers. The blue, orange, and yellow regions depict the ROA for the closed-loop system with the controllers $K_{LQR}$, $K_{max}$ and $K_{\mathcal{O}}$ respectively. 
Due to the complexity of computing the true ROA, we use a method to compute a close under-approximation of the true ROA. The true ROA of the controller $K_{max}$ is denoted by $R_\pi$, which is shown in green.

The trade-off between the LQR cost and the ROA for the three controllers as computed by our algorithm (presented in the next section) is shown in Table~\ref{T1}. 
The controller $K_{max}$ that optimizes the size of ROA
achieves an ROA which is $33.5\%$ higher than that of $K_{LQR}$, but at a cost of $6.1\%$ more LQR cost. 
The controller $K_{\mathcal{O}}$ synthesized by our algorithm by co-optimizing both LQR cost and ROA achieves $32.7\%$ better LQR cost than $K_{LQR}$ (close to $K_{max}$) with a significantly better LQR cost (almost $34\%$ better than $K_{max}$). For a given dynamical system, our goal in this paper is to develop a mechanism to synthesize $K_{\mathcal{O}}$ that achives an ROA close to that of $K_{max}$ and an LQR cost close to that of $K_{LQR}$.

In the next section, we present our algorithm for synthesizing $K_{\mathcal{O}}$ automatically.


\section{Algorithm}
\label{sec-algo}

In this section, we present our controller synthesis algorithm in detail.
Our algorithm is based on \emph{particle swarm optimization (PSO)}, an evolutionary computational method developed by  American scholars Kennedy and Eberhart in the early $90$s inspired by social behavior of fish schooling and bird flocking~\cite{KennedyE95}. The PSO algorithm solves a minimization problem of the following form:
  \begin{mydef}(\textit{Minimization Problem})
  Let $\mathcal{X} \subset \mathbb{R}^n $ be a convex search space then for a measurable function $\psi : \mathbb{R}^n \mapsto \mathbb{R}$ the minimization problem is to obtain a point $\mathbf{x}_0~\in~\mathcal{X}$  such that $\underset{\mathbf{x}\in\mathcal{X}}\min \ \psi(\mathbf{x})=\mathbf{x}_0$.
  \end{mydef}

The underlying idea for PSO is to seek for an optimal solution through particles or agents, whose trajectories are adjusted by a stochastic and a deterministic component. Each particle keeps track of its coordinates in the search space and they are influenced by its ``best'' achieved position called \textit{pbest} and the group's ``best'' position called \textit{gbest}. 
It employs an objective function, also known as \emph{fitness function}, to evaluate the effectiveness of each particle in an iteration, and through several iterations, it searches for the optimal solution. The iteration continues until convergence or for a pre-specified number of times.

Let $\bold{x}_k(m)$ be the position of the $k$-th particle in a set of $\aleph$ particles in the $m$-th iteration. 
Let $\bold{v}_k(m)$ be the updated velocity affected by a momentum factor $\omega$ that attracts the particles towards its earlier best positions $\bold{p}_k$ ($pbest$) and $\bold{g}$ ($gbest$) inside the whole swarm. The standard PSO algorithm uses the following equations to update  $\bold{v}_k(m)$ and $\bold{x}_k(m)$ for the $k$-th particle in the $(m+1)$-th iteration: 
\begin{align}
 \label{eq1} \bold{v}_k(m+1) = \omega \odot \bold{v}_k(m)& + \alpha \odot \textrm{rand}_1(\cdot) \odot (\bold{p}_k-\bold{x}_k(m)) \nonumber \\  & +\beta \odot \textrm{rand}_2(\cdot) \odot (\bold{g}-\bold{x}_k(m))  \\
 \label{eq2} \bold{x}_k(m+1) = \gamma \odot \bold{x}_k(m) & + \delta \odot \bold{v}_k(m+1)
\end{align}
where, $\odot$ signifies point-wise multiplication, $\alpha$ , $\beta$ yields the strength of attraction, $\textrm{rand}_1(\cdot) , \textrm{rand}_2(\cdot) \sim U(0,1)$ introduces the randomness for the good state space exploration.

We follow a deterministic version of the PSO algorithm as proposed by Trelea~\cite{Trelea03}. 
Though having the randomness in the PSO algorithm provides the guarantee to eventually converge to the global optimal solution, the progress rate of the algorithm is often very poor.
The deterministic algorithm proposed by Trelea~\cite{Trelea03} ensures fast exploration of the solution space, which is essential for the scalability of our synthesis procedure.

Following this algorithm, we deterministically choose the value for the two random numbers as follows:
\begin{align*}
    \textrm{rand}_1(\cdot) =\textrm{rand}_2(\cdot) = \frac{1}{2}
\end{align*}
and simplify Equation~\eqref{eq1} as follows:

\begin{align}
    \bold{v}_k(m+1) = \omega \odot \bold{v}_k(m) + \eta \odot(\rho_k - \bold{x}_k(m))
    \label{eq3}
\end{align}
where,
\begin{align*}
  \eta & =\frac{\alpha + \beta}{2}  \\
  \rho_k & = \frac{\alpha}{\alpha+\beta} \odot \bold{p}_k + \frac{\alpha}{\alpha+\beta} \odot \bold{g}
\end{align*}

Trelea~\cite{Trelea03} also shows that the parameters $\gamma$ and $\delta$ do not have any significant effect on the convergence of the algorithm. Following the suggestion in\cite{Trelea03}, we choose their values to be 1.
Thus, the algorithm depends on the two tuning parameters $\omega$ and $\eta$. 

In our methodology, the fitness function is  defined to map a candidate feedback controller in the search space to the size of its \emph{ROA} and its trajectory tracking performance in terms of \emph{LQR} cost. The controllers are the particle in the search space and in each step, we identify the \textit{gbest} particle according to the fitness function.

Initially a set of random coordinates of particles (controllers) with random positions and velocities are considered. Then the newer sets of coordinates get updated through PSO which run to obtain the global best particle ($gbest$) or configuration. 
The smallest LQR cost and the largest \emph{ROA} obtained locally so far for a particle is stored in the the $pbest$ variable which is then followed by the finding the global best solution $gbest$ among the set of all $pbest$.
Consequently, the optimal solution achieved through iterations. Algorithm~\ref{algo-synthesis} formally presents our synthesis algorithm. We now present the methodology to compute the LQR cost and ROA for a controller $K$ (the functions $\mathtt{lqr\_cost}$() and $\mathtt{roa}$() respectively).

\begin{algorithm2e}[t]

\DontPrintSemicolon
\textbf{Input:} particle positions and velocity \;
\textbf{Output: } updated positions and velocity\;
\BlankLine
\For{each particle $k$}
{
Initialize position $\bold{x}_{k}(0)$ and velocity $\bold{v}_{k}(0)$ uniformly at random within the permissible range\;
$\bold{p}_k \gets \bold{x}_{k}(0)$\;
}
$\bold{g} \gets \underset{\bold{p}_k}{\mathrm{argmin}} \ \mathtt{fitness}(\bold{p}_{k})$\;
\For{$m = 0$ to $max\_iter$}
{
\For{each particle $k$}
{
  Compute $\bold{v}_k(m+1)$ using Equation~\eqref{eq3}\;
  Compute $\bold{x}_k(m+1)$ using Equation~\eqref{eq2}\;
  $F_k \gets \mathtt{fitness}(\bold{x}_k(m+1))$ \;
  \If{$F_k < \mathtt{fitness}(\bold{p}_{k})$}
  {
    $\bold{p}_k \gets \bold{x}_k(m+1)$\;
    \If{$\mathtt{fitness}(\bold{p}_{k}) < \mathtt{fitness}(\bold{g})$}
    {
      $\bold{g} \gets \bold{p}_k$\;
    }
  }    
}
}

 \BlankLine
 \BlankLine
\SetKwProg{myproc}{Procedure}{}{}\label{p1}
\myproc{ $\mathtt{fitness}$($K$)} {
    $\mathtt{return}$ \ $\omega_1 \times \mathtt{lqr\_cost}(K) - \omega_2 \times \mathtt{roa} (K)$
}

\caption{Controller Synthesis Algorithm}
\label{algo-synthesis}
\end{algorithm2e}

\subsection{Computation of LQR cost}
Following a standard control-theoretic construction~\cite{Hespanha09}, the cost function (\ref{LQR_cost}) can be rewritten as $J_{LQR}=x[0]^\mathsf{T} P(K)x[0]$, where $u[r]=-Kx[r]$, and $P(K)\in\mathbb{R}^{n\times n}$ is a positive definite matrix that is the unique solution for $P$ to the Lyapunov equation:
\begin{equation}\label{lyapunov}
\left(A_\tau-B_\tau K\right)^TS\left(A_\tau-B_\tau K\right)-P+Q+K^TRK=0,
\end{equation}
where $K$ is a controller making $A_\tau-B_\tau K$ Hurwitz.\footnote{
We call the matrix $A_\tau-B_\tau K$ Hurwitz if its eigenvalues are inside the unit circle, centered at the origin.} 
Additionally, we have ${J_{LQR}}\leq\lambda_{\max}(P(K))\Vert{x[0]}\Vert^2$,

where $\lambda_{\max}(P(K))\in\mathbb{R}^+$ is the maximum eigenvalue of $P(K)$. 
Thus, $J_{LQR}$ can be minimized for all possible choices of initial conditions by minimizing the maximum eigenvalue of $P(K)$. Since $P$ is a symmetric positive definite matrix, its maximum eigenvalue is equal to its \emph{spectral norm} given by $\sqrt{\lambda_{\max}\left(P^\mathsf{T}P\right)}$~\cite{Meyer00}.

\subsection{Computation of the Region of Attraction}

For a given controller $K \in \mathcal{K}$, we can compute an under-approximation of $\mathcal{R}_K$ by using a \emph{Lyapunov function}. We have the following theorems in this concern.

\begin{theorem} (Lyapunov stability~\cite{Khalil02})
Suppose $f_K$ is locally Lipschitz continuous and has an equilibrium point at $x[0]=0$ and $\mathfrak{v}:\mathcal{S} \longmapsto \mathbb{R}$ be locally Lipschitz continuous on $\mathcal{S}$. If there exists a set $\Delta_{\mathfrak{v}} \subseteq \mathcal{S}$ containing $0$ on which $\mathfrak{v}$ is positive-definite and
\begin{align}
\label{c1}
  \mathfrak{v}((A_{\tau}-B_{\tau}K)x[r])< \mathfrak{v}(x[r]) ~~ \forall x[r] \in \mathcal{S}.
\end{align}
then $x[0]=0$ is an asymptotically stable equilibrium. In this case, $\mathfrak{v}$ is known as a
\emph{Lyapunov function} for the closed-loop dynamics $f_K$, and $\Delta_{\mathfrak{v}}$ is the \emph{Lyapunov decrease region} for $\mathfrak{v}$.
\label{T1}
\end{theorem}

Every level set $\lambda(c) = \{x[r]\ |\ \mathfrak{v}(x[r]) < c\}$, 
$c\in \mathbb{R}^+$ such that $\lambda(c) \subseteq \Delta_{\mathfrak{v}}$ is an ROA for $f_K$ and $x[0]=0$.

Instead of searching a pertinent Lyapunov candidate based on the computational methods that constrain the search space to a very specific class of functions, Berkenkamp et al~\cite{RichardsB018} proposes a technique to construct the  Lyapunov candidate $v_\theta(x)$ as an inner product of  feed-forward neural networks as $v_\theta(x) = \phi_{\theta}(x)^\mathsf{T}\phi_{\theta}(x)$. This neural network function  $\phi_{\theta}(x)$, called the \emph{Lyapunov Neural Network} consists of a sequence of layers. 
Each output layer is parameterized by a suitable weight matrix that yields an input to a fixed element-wise activation function. 
Berkenkamp et al~\cite{RichardsB018} provide sufficient conditions on the weight matrix and the activation function of the neural network $\phi_{\theta}(x)$  to be both positive definite and locally Lipschitz continuous, essential conditions for $\phi_{\theta}(x)$ to become a Lyapunov function candidate (see Theorem 2 in~\cite{RichardsB018}). 

 \begin{theorem}(Lyapunov Neural Network)~\cite{RichardsB018}\\
 Consider $\mathfrak{v}_{\theta}(x)= \varphi_{\theta}(x)^\mathsf{T}\varphi_{\theta}(x)$ as a Lyapunov candidate function, where $\varphi_{\theta}(x)$ is a feed-forward neural network. Suppose, for each layer $\ell$ in $\varphi_{\theta}(x)$, the activation function $\varphi_{\theta}(x)$ and the weight matrix $\mathbf{W}_{\ell} \in \mathbb{R}^{n_{\ell}\times n_{\ell-1}}$
  each have a trivial nullspace. Then $\varphi_{\theta}$
 has a trivial nullspace as well, and $\mathfrak{v}_{\theta}$ is positive-definite with $\mathfrak{v}_{\theta}(\mathbf{0})=0$ and $\mathfrak{v}_{\theta}(x)>0~,\forall ~x \in \mathcal{S} \sim \{\mathbf{0}\}$.
 Moreover, if $\mathfrak{v}_{\theta}$ is Lipschitz continuous for each layer $\ell$, then $\mathfrak{v}_{\theta}$ is also locally Lipschitz continuous.
 \end{theorem}
 
Theorem~\ref{T1} provides a sufficient condition~(\ref{c1}) to obtain the set of safe states, however it is extremely difficult to verify this on a continuous subset $\Delta_{\mathfrak{v}}$. 
The methodology proposed by Berkenkamp et al.~\cite{RichardsB018} learns the Lyapunov neural network   
by checking the \emph{tightened safety certificate} 
\begin{align}
\label{c2}
 \mathfrak{v}_{\theta}((A_{\tau} - B_{\tau}K)x[r]))- \mathfrak{v}_{\theta}(x[r]) +L_\mathfrak{v} \mu < 0   
\end{align}
at a set of discrete state points covering the original set $\mathcal{S}$. Here, $L_{\mathfrak{v}} \in \mathbb{R}^{+}$ is a Lipschitz constant and $\mu \in \mathbb{R}^{+}$ is the measure of how densely those points cover the continuous state space.
 Following the method of estimating \emph{ROA} to be as much of  $\mathcal{R}_{K}$ as possible for a given controller $K$, we develop an algorithm to scrutinize the best possible policies based on the performance measure of \emph{LQR} cost and \emph{ROA}. We implement the following optimization technique to learn such policy through sequential iteration.
We apply condition~(\ref{c2}) at a finite set of points covering $\mathcal{S}$ in ascending order of the values of $\mathfrak{v}_{\theta}(\mathbf{x})$.
The neural network finally converges to represent a Lyapunov function whose largest level set closely represnt the true ROA for the system $f_K$.

Convergence isn't guaranteed and it may happens that with increased iteration the volume may shrink instead growing monotonically towards $\mathcal{R}_{\pi}$. However there is a trade-off.

\section{Experiments}
\label{sec-experiments}

In this section, we demonstrate the effectiveness of our proposed methodology through experimental results on various example applications.

\subsection{Implementation and Experimental Setup}
We implement our PSO based controller synthesis algorithm in the Matlab environment. To compute the ROA for a fixed control policy, we use the Python implementation available in~\cite{safe_learning}, which we adapt suitably and interface with our Matlab implementation. The implementation for computing ROA uses the Stocastic Gradient Discent (SGD) optimization method 
in the TensorFlow framework~\cite{AbadiBCCDDDGIIK16} to train the Lyapunov Neural Networks corresponding to different controllers.

We discretize the linearized dynamics with a time step of $\tau = 0.01\si{\second}$.
We set the maximum number of iterations $max\_iter$ in our PSO based algorithm to $15,000$. In our implementation, we have an additional termination condition. If the value of  $gbest$ does not change in the last $100$ iterations, we terminated the execution of the algorithm. 
Each data point presented in Section~\ref{sec-results} is the average of the results obtained in $30$ runs.
Following a large number of simulation experiments, we choose two sets of parameters for the PSO algorithm and use them with equal probability in our experiments: $\langle \omega = 0.7, \eta= 1.6\rangle$ and $\langle \omega= 0.33, \eta=2.35 \rangle$.
We have carried out our experiments on a Linux machine with Intel Core i$7$-$8700$ Hexa-core CPU with clock speed at max $4600$MHz and $32$GB RAM.

\subsection{Examples}

We have carried out our experiments on two variants of inverted pendulum, a vehicle steering model, and an aircraft pitch control model. The dynamics of the inverted pendulum has been provided in Section~\ref{sec-example}. Below we provide the details of the other two models.

\begin{table*}
\small
\centering

    \begin{tabular}{|c|c|c|c c|c c|c|}
                        \hline   
                        \multicolumn{1}{|c|}{\textbf{\texttt{Benchmark}}} &         
                        \multicolumn{1}{c|}{\makecell{\textbf{\texttt{Range }}\\$[\bold{x}_{\min},\bold{x}_{\max}]$}} &  
                        \multicolumn{1}{c|}{\makecell{\textbf{\texttt{No of }} \\\textbf{\texttt{Particles}}\\ \texttt{$(\aleph)$}}} &                       
                        \multicolumn{2}{c|}{\makecell{\textbf{ $\%$  \texttt{Increase}}\\ \textbf{\texttt{in LQR cost}}}} & \multicolumn{2}{c|}{\makecell{\textbf{ $\%$  \texttt{Increase}}\\ \textbf{\texttt{in ROA}}}}&
                        
                        \multicolumn{1}{c|}{\makecell{\textbf{ \texttt{Expected Time}}\\ \textbf{\texttt{ Costs (approx) }}}}\\
                        \cline{4-7}
                         & & & 
                         $K_{\mathcal{O}}$ & $K_{max}$ &
                         $K_{\mathcal{O}}$ & $K_{max}$ &\\
                         \hline
                       \makecell{\texttt{Pendulum}\\ \texttt{A}} & \makecell{$[-10,10]$ \\ $\times [-5,5]$} &\makecell{$10$\\$15$\\$20$\\$25$\\$30$} &  \makecell{$1.5$ \\ $2.2$ \\ $3.67$\\$4.02$ \\ $4.89$}  & \makecell{$2.25$ \\ $3.5$ \\ $4.7$\\ $6.1$\\ $6.77$} & \makecell{$15.8$ \\ $18.3$ \\ $23.2$ \\$32.7$ \\ $33.65$} & \makecell{$16.91$ \\ $20.1$ \\ $25.78$ \\ $33.5$ \\ $34.87$} & \makecell{$45$ mins $\sim$ $2$ hrs}\\
                     \hline
                     \makecell{\texttt{Pendulum}\\ \texttt{B}} & \makecell{$[5,20]$\\ $\times [-2,12]$} & \makecell{$10$\\$15$\\$20$\\$25$\\$30$} & \makecell{$0.8$ \\ $1.5$ \\ $2.01$ \\ $2.65$ \\$3.13$}  &\makecell{$1.2$ \\ $2.12$ \\ $2.89$ \\ $3.31$ \\$3.94$}  & \makecell{$5.7$ \\ $10.125$ \\ $12.62$ \\ $17.43$ \\ $19.75$} & \makecell{$8.41$ \\ $13.94$ \\ $15.89$ \\ $20.1$ \\ $22.3$} & \makecell{$45$ mins $\sim$ $2$ hrs }\\
                        \hline
                     \makecell{\texttt{Vehicle} \\\texttt{Steering}} & \makecell{$[0,17]$ \\ $\times [0,11]$} &\makecell{$10$\\$15$\\$20$\\$25$\\$30$}   & \makecell{$2.19$ \\ $3.63$ \\ $4.86$ \\ $5.61$ \\ $6.91$} & \makecell{$3.41$ \\ $4.32$ \\ $5.27$ \\ $6.65$ \\ $8.13$}  & \makecell{$12.29$ \\ $14.03$ \\ $24.72$ \\ $26.84$ \\ $31.05$} & \makecell{$14.47$ \\ $16.31$ \\ $25.75$ \\$28.50$ \\ $33.73$} & \makecell{$1$ hr $\sim$ $3$ hrs} \\
                     \hline
                     \makecell{\texttt{Aircraft} \\\texttt{Pitch}} 
                     & \makecell{$[-1,5]$\\ $\times [10,100]$ \\ $\times [0,7]$} 
                     &\makecell{$10$\\$15$\\$20$\\$25$\\$30$}   & \makecell{$3.24$ \\ $5.04$ \\ $5.90$ \\ $6.50$ \\ $7.90$} & \makecell{$4.89$ \\ $6.16$ \\ $7.23$ \\ $8.57$ \\ $10.12$}  & \makecell{$9.55$ \\ $18.72$ \\ $20.20$ \\ $21.50$ \\ $23.10$  } & \makecell{$11.01$ \\ $20.51$ \\ $22.76$ \\ $24.28$ \\ $25.23$} & \makecell{$2$ hrs $\sim$ $7$ hrs}\\ 
                     \hline
                \end{tabular}
                \caption{Algorithm Performance}\label{T2}
              \end{table*}

\smallskip
\noindent
\textit{Vehicle Steering}~\cite{astrom2010feedback}. An important subsystem of a vehicle operation is a steering system which helps in turning the wheel in different driving conditions. 
If the driver attempts to enter a curve too fast, she may lose control on vehicle and ultimately ends up in fatal crashes. 
Vehicle crashes due to dangerous turns can be reduced if the \emph{ROA} of controller is known and suitable warning signal can be issued whenever a violation is about to happen. Often \emph{LQR} controllers are employed to improve the performance of a steering system and to obtain an optimal cost controller but it may have low \emph{ROA}. Many articles on \emph{Electronic Power Steering} system (EPS) have been published so far ~\cite{astrom2010feedback,Nise07,ChituLHWK11,XiangX06}. However, our major contribution lies in synthesizing an  controller by co-optimizing both the \emph{ROA} and the \emph{LQR} cost. The nonlinear equations of motion of a simple vehicle can be expressed in the following form:
\begin{align*}
 \dot{x} & = v \cos(\alpha(\delta)+\theta)\\
 \dot{y} & = v \sin(\alpha(\delta)+\theta)\\
 \dot{\theta} & = \frac{v_0}{\mathfrak{b}} \tan \delta
\end{align*}
where $v_0$ is the velocity of the rear wheel, $x$, $y$ and $\theta$ are the position and orientation of the center of gravity of the vehicle, $\mathfrak{b}$ is the distance between the front and rear wheels and $\delta$ is the angle of the front wheel. 
The function $\alpha(\delta)$ is the angle between the velocity vector and the vehicle’s length axis.
In our experiments, we consider such model by approximating the motion of the front and rear pairs of wheels by a single front wheel and a single rear wheel, to obtain  an abstraction of a bicycle model. 

\smallskip
\noindent
\textit{Aircraft Pitch Control}\cite{Bowyer92}.

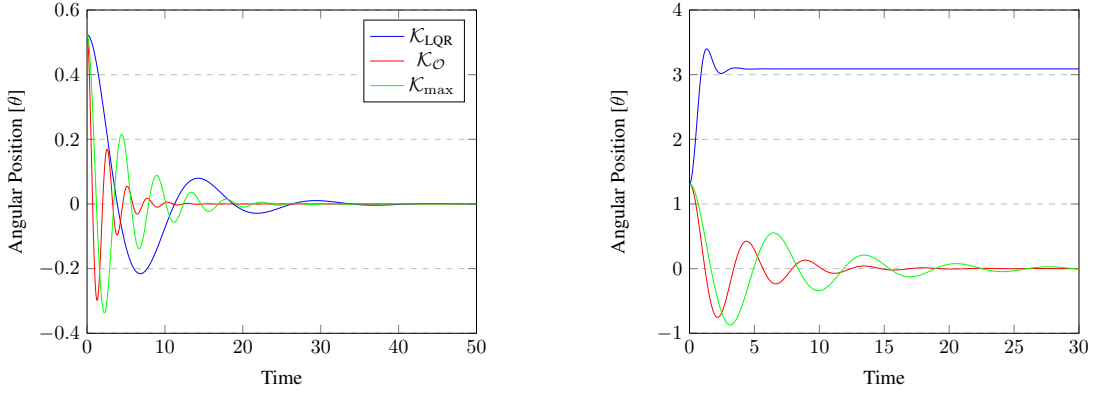
\begin{figure*}
\centering
\resizebox{0.4\textwidth}{!}{%

\begin{tikzpicture}
\begin{axis}[
    xlabel={Time},
    ylabel={Angular Position [$\theta$]},
    xmin=0, xmax=50,
    ymin=-0.4, ymax=0.6,
    xtick={0,10,20,30,40,50},
    ytick={-0.4,-0.2,0,0.2,0.4,0.6},
    legend pos=north east,
    ymajorgrids=true,
    grid style=dashed,
]

\addplot[
    color=blue,
    ]
    table [x=x, y=a, col sep=comma] {T20.csv};
    \addplot[
    color=red,
    ]
    table [x=x, y=b, col sep=comma] {T20.csv};
    \addplot[
    color=green,
    ]
    table [x=x, y=c, col sep=comma] {T20.csv};
    \addlegendentry{$\mathcal{K}_{\text{LQR}}$ }
    \addlegendentry{$\mathcal{K}_{\mathcal{O}}$}
    \addlegendentry{$\mathcal{K}_{\max}$}
    
\end{axis}
\end{tikzpicture}
}%
\qquad
\qquad
\resizebox{0.4\textwidth}{!}{%
\begin{tikzpicture}
\begin{axis}[
    xlabel={Time},
    ylabel={Angular Position [$\theta$]},
    xmin=0, xmax=30,
    ymin=-1, ymax=4,
    xtick={0,5,10,15,20,25,30},
    ytick={-1,0,1,2,3,4},
    ymajorgrids=true,
    grid style=dashed,
]

\addplot[
    color=blue,
    ]
    table [x=x, y=a, col sep=comma] {Test20.csv};
    \addplot[
    color=red,
    ]
    table [x=x, y=b, col sep=comma] {Test20.csv};
    \addplot[
    color=green,
    ]
    table [x=x, y=c, col sep=comma] {Test20.csv};
    \end{axis}
\end{tikzpicture}
}%
\caption{Comparison of the performance of the controllers when the pendulum is driven from $c=\pi/6$ (left) and $c=5\pi/12$ (right).}
\label{1}
\end{figure*}

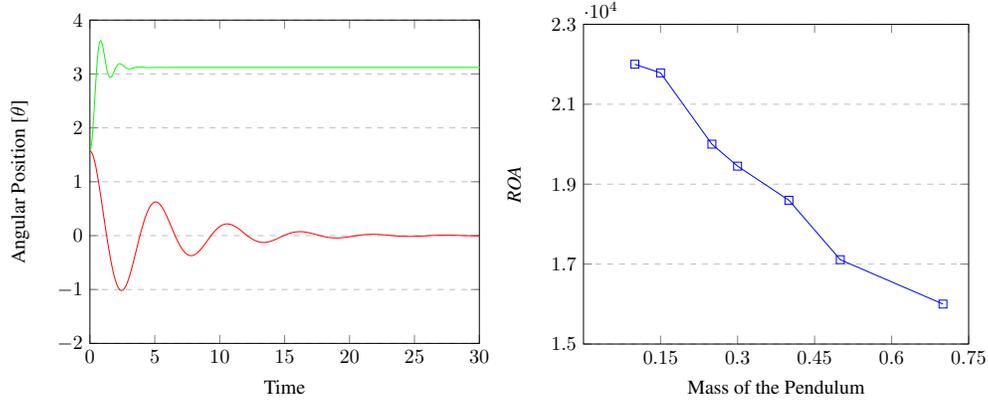
\begin{figure}
\centering
 \resizebox{0.4\textwidth}{!}{%
 \begin{tikzpicture}
 \begin{axis}[
  xlabel={Time},
     ylabel={Angular Position [$\theta$]},
     xmin=0, xmax=30,
     ymin=-2, ymax=4,
     xtick={0,5,10,15,20,25,30},
    ytick={-2,-1,0,1,2,3,4},
     ymajorgrids=true,
     grid style=dashed,
 ]

 \addplot[
     color=green,
     ]
     table [x=x, y=a, col sep=comma] {F21.csv};
     \addplot[
     color=red,
     ]
     table [x=x, y=b, col sep=comma] {F21.csv};
    
 \end{axis}
 \end{tikzpicture}
 }%
\resizebox{0.4\textwidth}{!}{%
\begin{tikzpicture}
\begin{axis}[
    xlabel={Mass of the Pendulum},
    ylabel={\emph{ROA}},
    xmin=0, xmax=0.75,
    ymin=15000, ymax=23000,
    xtick={0.15,0.30,0.45,0.6,0.75},
    ytick={15000,17000,19000,21000,23000,25000},
    legend pos=north west,
    ymajorgrids=true,
    grid style=dashed,
]

\addplot[
    color=blue,
    mark=square,
    ]
    coordinates {
    (0.1,22000) (0.15,21784) (0.25,20001) (0.3,19451) (0.4,18597) (0.5,17110) (0.7,16000)
    };
    
\end{axis}
\end{tikzpicture}
}%
\caption{Stability comparison of rest two controllers when  $c=\pi/2$ (left) and graph of \emph{ROA} w.r.t increasing mass (right).}
\label{fig2}
\end{figure}

\begin{figure*}[h!]
\centering
\resizebox{0.4\textwidth}{!}{%
\begin{tikzpicture}
\begin{axis}[
  xlabel={Number of discrete states along each dimension},
    ylabel={Computation Time},
    xmin=200, xmax=260,
    ymin=20, ymax=400,
    xtick={200,220,240,260,280},
    ytick={0,50,100,150,200,250,300},
    legend pos=north west,
    ymajorgrids=true,
    grid style=dashed,
]

\addplot[
    color=blue,
    mark=square,
    ]
    coordinates {
    (200,30) (220,32) (240,35) (250,39) (260,55)
    };
    \addplot[
    color=red,
    mark=square,
    ]
    coordinates {
    (200,33) (220,36) (240,43) (250,49) (260,66)
    };
    \addplot[
    color=green,
    mark=square,
    ]
    coordinates {
    (200,121) (220,132) (240,156) (250,250) (260,310)
    };
    \addlegendentry{Inv. Pendulum }
    \addlegendentry{Veh. Steering}
    \addlegendentry{Aircraft Pitch}

\end{axis}
\end{tikzpicture}
}%
\qquad
\qquad
\resizebox{0.4\textwidth}{!}{%
\begin{tikzpicture}
\begin{axis}[
    xlabel={Number of discrete states along each dimension},
    ylabel={\emph{ROA}},
    xmin=200, xmax=260,
    ymin=6000, ymax=22000,
    xtick={200,220,240,260,280},
    ytick={6500,9500,12500,15500,18500,23500},
    ymajorgrids=true,
    grid style=dashed,
]

\addplot[
    color=blue,
    mark=square,
    ]
    coordinates {
    (200,12822) (220,15400) (240,18497) (250,19728) (260,21418)
    };
    \addplot[
    color=red,
    mark=square,
    ]
    coordinates {
    (200,9782) (220,11468) (240,12456) (250,13289) (260,15212)
    };
    \addplot[
    color=green,
    mark=square,
    ]
    coordinates {
    (200,6689) (220,7654) (240,9872) (250,10235) (260,10876)
    };
    
\end{axis}
\end{tikzpicture}
}%
\caption{Computation time in seconds (left) and \emph{ROA} (right) vs number of discrete states in the state space. }
\label{fig3}
\end{figure*}
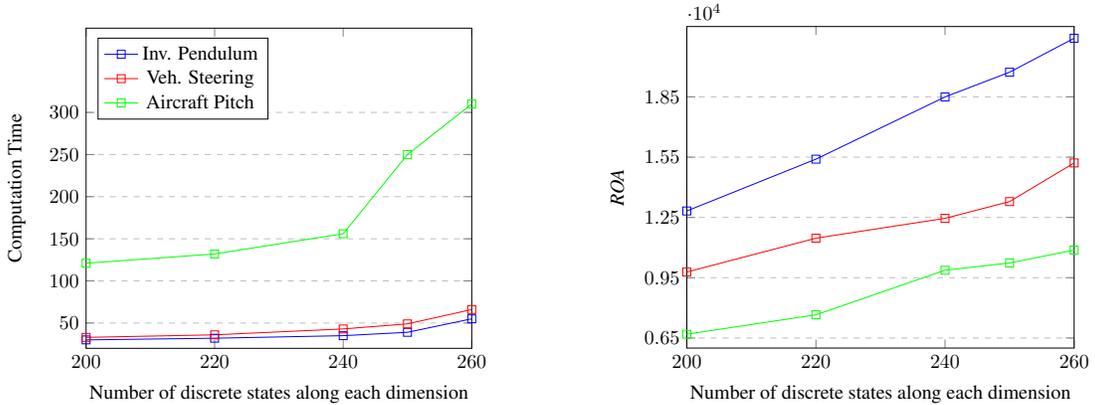

The  stability of an aircraft is governed mainly by three factors: Center of Gravity (CG), Center of Pressure (CP), and the design of elevator. The further the CG is, the more stable the aircraft is with respect to pitching axis. However, far-forward CG position is also uncontrollable, so there is a trade-off. Because, if CG is aft to CP then the longitudinal stability might decrease, which isn't  a favourable situation. From the viewpoint of stability, CG should always coincide with CP. However, CP is not static and can move depending on the angle of incidence of the wings. The role of elevator is to control the pitching rotations of the aircraft. As a result, larger amount of engine thrust is required to maintain the static pitching stability of an aircraft. The longitudinal equations of motion  along $x$, $y$, and $z$ axes are given as follows:
\begin{align*}
F_x & - mg \sin \theta  = m (\dot{v}_x-\omega_zv_y +\omega_yv_z)\\
M & = J_y \dot{\omega}_y +J_{xz} (\omega_x^2-\omega_z^2)+\omega_x \omega_z(J_x-J_z)\\
F_z & + mg \cos \theta \cos \phi = m(\dot{v}_z +\omega_x v_y -\omega_y v_x)
\end{align*}
where $F_x , F_z$ are the aerodynamic and propulsive forces acting along $x$ and $z$ directions, $\overrightarrow{v}=(v_x,v_y,v_z)$ is linear velocity,  $\overrightarrow{\omega}=(\omega_x,\omega_y,\omega_z)$ is angular velocity, $\overrightarrow{J}= (J_x,J_y,J_z)$ represents Moment of Inertia and $\theta, \phi$ are the pitch and roll angle respectively.
In our experiments, we consider an aircraft model in steady-cruise at constant altitude and velocity so that the thrusts, weight and lift forces can balance each other. 

\subsection{Results}
\label{sec-results}

\subsubsection{LQR cost vs. ROA trade-off}

We present our main results in Table~\ref{T2}. We run our PSO based synthesis algorithm for Swarm sizes of $\aleph= 10,15,20,25,30$ particles and suitabley chosen ranges $[x_{\min},x_{\max}]$ for the particle values as shown in Table~\ref{T2}. 
 Our results are consistent for all the systems and we can make the following conclusions. 
 The ROA increases monotonically with the increase in the number of particles.
 For a chosen number of particles, the ROA obtained for $K_{\mathcal{O}}$ is close to that of $K_{max}$, though the LQR cost of $K_{\mathcal{O}}$ is significantly less than that of $K_{max}$. Table~\ref{T2} also presents the range of the synthesis time for each system, where the synthesis time increases linearly with number of particles. 

 \subsubsection{Benefit of larger ROA}
We have created a Simulink model to estimate the ROA for different controllers through simulation. 

The simulation results for a pendulum model with different controllers are shown in Figure~\ref{1}.
From Figure~\ref{1}, we observe that all the three controllers $K_{LQR}$, $K_{max}$, and $K_{\mathcal{O}}$ succeed to bring the pendulum from an angle of $\pi/6$ degree to the upright position. However, though $K_{max}$, and $K_{\mathcal{O}}$ can bring the pendulum from an angle of $5\pi/12$ to the upright position, the controller $K_{LQR}$ fails to do so. This observation clearly establishes that both $K_{max}$, and $K_{\mathcal{O}}$ provide better ROA than $K_{LQR}$, which was our main goal for controller synthesis.

 \subsubsection{ROA for pendulums with different masses}
We have carried out experiments with pendulums with a fixed length and different masses between $0.1\si{\kilogram}$\ -\ $0.7\si{\kilogram}$. Intuitively, one could expect that for a fixed limit on possible actuation, it is more difficult to bring a pendulum with larger mass to the vertical position from a state further from the unstable equilibrium point. That is, a pendulum with a larger mass has a smaller ROA. This intuition is also evident from our experimental results  presented in Figure~\ref{fig2}. The figure shows that the ROA of the synthesized controller decreases monotonically with the increase in the mass.

\subsubsection{Effect of state space discretization}
An important parameter of our algorithm is $\mu$ which decides how finely we discretize the continuous state space.
Figure~\ref{fig3} shows how the computation time of the ROA for different dynamical systems and the size of the ROA vary with the increase in the value of the discretization factor $\mu$.
The figure shows that While having a finer discretization helps us in capturing the actual ROA of the closed loop system more accurately, it also leads to a blowup in the computation time.

\section{Related Work}
\label{sec-related}
In this section, we will briefly explore some of the major works that are closely related to ours.

Safe-learning is an active area of research that has been drawn prominent attention from both the researchers in machine learning and control communities~\cite{AswaniGST13,BerkenkampKS16,AkametaluKFZGT14}. Discrete \emph{Markov Decision Process}, \emph{Model Predictive Control} scheme these are few areas that has considered the existence of feasible return trajectories to a safe region of the state space with high probability. Nevertheless for a non-linear dynamical system \emph{Lyapunov functions} are the most convenient tools for safety certification and hence \emph{ROA} estimations~\cite{VannelliV85,SilvaT05}.Even though searching such function analytically is not a straight forward task but can be identified efficiently via a  semi definite program~\cite{Parrilo00,BV2014} , or using SOS polynomial methods~\cite{HenrionK14}. Some other methods to obtain  \emph{ROA} includes volume over system trajectories, sampling based approaches~\cite{BobitiL16} and so on.

\section{Conclusion and Future Work}
\label{sec-conclusion}
We have developed a novel method for synthesizing control policies for general nonlinear dynamical systems. 
Our work borrows insights from recent advances in estimating ROA for a nonlinear dynamical system, resulting in an algorithm that can synthesize control policies by minimizing the LQR cost as well as expanding the  ROA, which is essential for energy-efficient, safe, and high-performance operations of life-critical and mission-critical embedded applications.

An interesting area that we want to broadly explore in the future is to initiate the learning process and update the weights of the neural network through deep reinforcement learning (RL) technique. Once the deep neural network for RL is initialized, a reward function that captures both the trajectory tracking capability and the size of the ROA will be used to improve the feedback controller. We also plan to synthesize feedback controllers for complex dynamical systems like quadcopters. Our final goal would be to fly a UAV by using the feedback controllers synthesized through the proposed technique and evaluate the efficacy of the synthesized controller under various disturbance conditions.

\bibliographystyle{unsrt}  
\bibliography{references}

\end{document}